\documentstyle[psfig]{article}

\def\dm0{d{\tilde \mu}_{0}}

\newcommand\E{{\tilde{E}}}
\newcommand\A{A_a}
\newcommand\maps{\colon}

\def\ut#1{\rlap{\lower1ex\hbox{$\sim$}}{#1}}
\def\IP#1#2{\langle\,#1\,|\,#2\,\rangle}
\def\Tr{\mathop {\rm Tr}}

\newcommand{\B}{\cal B}
\newtheorem{theorem}{Theorem}
\newcommand\tensor{\otimes}
\renewcommand\a{\overline{\cal A}}
\newcommand\f{\overline{\cal F}}
\newcommand\g{{\cal G}}
\newcommand\ag{\a/\g}
\newcommand\hfull{L^2(\a \times \f)}
\newcommand\hinv{L^2((\a \times \f)/\g)}
\newcommand\hdiff{{\cal H}_{\rm diff}}

\newcommand\diff{{\rm diff}}

\newcommand{\Fun}{{\rm Fun}}
\newcommand{\SU}{{\rm SU}}
\newcommand{\U}{{\rm U}}

\newcommand{\Ad}{{\rm Ad}}
\newcommand{\iso}{\cong}

\newcommand{\pic}[5]{\raisebox{#3pt}{
\hspace{#4pt}\psfig{file=#1.ps,height=#2pt,silent=}\hspace{#5pt}}}
\newcommand{\kd}[1]{\mathchoice{
\pic{#1}{12}{5}{-1}{2}}{
\pic{#1}{13}{-6}{-3}{2}}{
\pic{#1}{9}{-2}{-3}{1}}{
\pic{#1}{7}{-1}{-1}{0}}}

\begin{document}

      \begin{center}
      {\large {\bf Quantization of Diffeomorphism-Invariant \\ 
      Theories with Fermions}\\}
      \vspace{0.5cm}
      {\em John C.\ Baez\\}
      \vspace{0.3cm}
      {Department of Mathematics,  University of California \\
      Riverside, California 92521, USA \\}
      \vspace{0.3cm}
      {\em Kirill V.\ Krasnov\\}
      \vspace{0.3cm}
      {Center for Gravitational Physics and Geometry, Pennsylvania State 
      University \\ University Park, PA 16802, USA\\
      and\\
      Bogolyubov Institute for Theoretical Physics, Kiev 143, Ukraine\\ }
      \vspace{0.3cm}
      {\small email: baez@math.ucr.edu, krasnov@phys.psu.edu\\}
      \vspace{0.3cm}
      {\small \today\\}
      \end{center}

\begin{abstract}
We extend ideas developed for the loop representation of quantum
gravity to diffeomorphism-invariant gauge theories 
coupled to fermions.  Let $P \to \Sigma$ be a
principal $G$-bundle over space and
let $F$ be a vector bundle associated to $P$ whose fiber is
a sum of continuous unitary irreducible representations of 
the compact connected gauge group $G$,
each representation appearing together with its dual.  We
consider theories whose classical configuration space is ${\cal 
A} \times {\cal F}$, where ${\cal A}$ is the space of connections on 
$P$ and $\cal F$ is the space of sections of $F$, regarded as a 
collection of Grassmann-valued fermionic fields.   We construct
the `quantum configuration space $\a \times \f$ as a completion
of ${\cal A} \times {\cal F}$.  Using this we construct a Hilbert space 
$L^2(\a \times \f)$ for the quantum theory on which all automorphisms of 
$P$ act as unitary operators, and determine an explicit `spin network basis' 
of the subspace $\hinv$ consisting of gauge-invariant states.  
We represent observables constructed from holonomies of the connection 
along paths together with fermionic fields and their conjugate momenta as 
operators on $\hinv$.  We also construct a Hilbert space $\cal H_\diff$
of diffeomorphism-invariant states using the group averaging procedure
of Ashtekar, Lewandowski, Marolf, Mour\~ao and Thiemann.  
\end{abstract}

\eject
\section{Introduction}

In this paper we address the problem of the quantization of a wide
class of diffeomorphism-invariant theories.  We consider theories
whose degrees of freedom are described in the Hamiltonian framework by
a connection on some principal $G$-bundle on space together with a set of
fermionic fields on space taking values in representations of the
compact connected gauge group $G$.  Theories of this class can be
thought of as giving a diffeomorphism-invariant description of
fermionic matter interacting by means of gauge fields.  One of the
main motivations for considering such theories is that general
relativity can be reformulated as a diffeomorphism-invariant theory having
as canonically conjugate variables a connection on space and a densitized
triad field \cite{NewVar}.  This means that the gravitational force
coupled to Yang-Mills gauge fields and fermionic matter fits into our
framework.

Our approach builds upon existing work on the loop representation of
quantum gravity and other diffeomorphism-invariant theories of
connections \cite{ALMMT}. The loop quantization of theories of connections
is based on an assumption that Wilson loop functionals of connection
become well-defined quantum operators.  Similarly, our approach is
based on an assumption that certain fermionic analogs of Wilson
observables become well-defined operators.  An important feature of
our scheme is that we do not introduce closed Wilson loops as
fundamental observables, as typical in the loop representation of a
pure gauge theory.  Instead, we use open Wilson paths with fermions or
their conjugate momenta at the endpoints, which we call `fermionic 
path observables'.  Closed loops arise as secondary quantities. 

The basic strategy of the paper can be summarized as follows.  We
introduce a set of path observables on the classical phase space of
our theory that forms a closed algebra with respect to Poisson
brackets.  The main idea then is to construct a `kinematical' Hilbert
space for the theory and represent our algebra by operators on this
Hilbert space.  This is not the space of physical states, because the
constraints --- gauge-invariance, spatial diffeomorphism-invariance,
and the Hamiltonian constraint --- have not yet been implemented.  The
quantization procedure we adopt satisfies some natural requirements;
for example, gauge-invariant classical quantities become operators
that preserve the gauge-invariant subspace of the kinematical Hilbert
space, `real' quantities become self-adjoint operators, and so on.  It
turns out, however, that representation of our algebra is, in general,
reducible even on the subspace of gauge-invariant states.  

An important issue in quantizing diffeomorphism-invariant gauge
theories coupled to fermions is that of respecting the classical
reality conditions on the quantum level.  In field theories on
Minkowski spacetime the operator $\dagger$ of Hermitian conjugation of
fermionic fields sends configuration fermionic fields into their
canonically conjugate momenta.  However, in diffeomorphism-invariant
theories the $\dagger$ operation is more complicated.  Consider for example
a theory in which the only gauge field is the gravitational field,
described by the connection field $A_a$ canonically conjugate to
the densitized triad field \cite{NewVar}.  The
configuration and momentum fermionic fields have different density
weights, and the action of $\dagger$ operator involves the square root
of the determinant of the metric:
\begin{equation}
\tilde{\pi}^\dagger = -i\sigma\xi.
\label{dagger}
\end{equation}
Here $\sigma$ is the square root of the determinant of the metric, 
defined in terms of the densitized triad field.

In the quantum theory, one wants to impose this `reality condition' in
an appropriate form.  However, in trying to impose the reality
condition (\ref{dagger}) in a naive way one encounters the following
problem.  In general relativity the triad field is a dynamical
variable.  Thus, in the quantum theory the quantity $\sigma$
constructed from the triad field becomes an operator.  Naively one
might require that the adjoint of the operator $\hat{\tilde \pi}$
corresponding to the fermionic momentum be given by:
\[
(\hat{\tilde{\pi}})^\dagger = -i\hat{\sigma}\hat{\xi}.
\]
To illustrate the problems arising from this naive definition,
consider the commutator 
\[
\left [ \int d^nx \,\hat{\tilde{\pi}}(x) f(x) , 
\int d^nx \Tr(\hat{A}_a(x) g^a(x))
\right ]
\]
of operators $\hat{\tilde{\pi}}$ and $\hat{A}_a$ smeared with test functions. 
For simplicity of the argument that follows let us consider the
case of a real connection $A_a$, corresponding to Riemannian general
relativity.  One might expect this commutator to vanish, 
since the corresponding Poisson bracket is zero in the classical theory.  
However, if we apply the $\dagger$ operation to this commutator, we obtain
\[
\left [ \int d^nx (-i\,\hat{\sigma})\hat{\xi}(x) , 
\int d^nx \Tr(\hat{A}_a(x) g^a(x))\right ] 
\]
which appears to be nonzero, since the operator $\hat{\sigma}$ does 
not commute with $\hat{A}_a$.  Thus, the $\dagger$ operation defined 
in this naive way seems to lead to inconsistencies.

Our strategy for dealing with this issue is as follows.  One role of
the $\dagger$ operation on functions on the classical phase space is
to say which of these functions are real.  Therefore, a possible way
to impose the classical reality conditions on the quantum level is to
require that classical real quantities are represented by self-adjoint
operators at the quantum level.  One then has a satisfactory quantum
theory if a sufficient number of physically interesting real
observables are represented by self-adjoint operators.  The
quantization scheme adopted in this paper respects the classical
reality conditions in the sense that a certain set of real observables
is represented by self-adjoint operators. Note, however, that in this
paper only a rather small set of real observables is treated.  For
instance, the quantization of the Hamiltonian constraint, which is
itself a real functional on the phase space of the theory (in its
`real' formulation used in \cite{Thiemann}), is not considered in this
paper.

The organization of this paper is as follows.  In Sec.\ \ref{sec:2} we
start by reviewing the main results of the loop representation
approach to quantizing diffeomorphism-invariant theories of
connections.  In Sec.\ \ref{sec:3} we specify the class of theories
which are of interest for us here and give some examples of theories
belonging to this class.  In Sec.\ \ref{sec:4} we describe the path
observables which are the basic building blocks of our quantum theory.
In Sec.\ \ref{sec:5} we describe the classical configuration space and
a certain completion of it which we call the `quantum configuration
space'.  We then construct the kinematical Hilbert space as a space of
functions on the quantum configuration space.  In Sec.\ \ref{sec:6} we
construct the Hilbert space of gauge-invariant states, and describe a
basis of this space given by `open spin networks' with ends labelled
by fermionic fields.  In Sec.\ \ref{sec:7} we define operators
representing the path observables.  The issue of
diffeomorphism-invariant states is discussed in Sec.\ \ref{sec:8}.  We
conclude with a discussion of the results obtained, focusing on the
issue of superselection sectors.

Despite its great importance, we do not discuss dynamics in this
paper, as the correct treatment of the Hamiltonian constraint is still
a matter of controversy in the loop representation of pure gravity,
despite recent progress on this front \cite{Thiemann}.  We hope,
however, that our work will serve as a basis for future work on
dynamical aspects of the quantization of diffeomorphism-invariant
gauge theories coupled to fermions.

\section{Quantization of diffeomorphism-invariant \hfill \break theories of 
connections}
\label{sec:2}

This section reviews the main steps of quantization of 
diffeomorphism-invariant theories of connections.  
See references \cite{AL,ALMMT,B} for more details.

In the Hamiltonian formalism, the kinematical phase space of such a
theory consists of pairs $(A,{\tilde E})$ satisfying suitable
regularity conditions.  Here $A$ is a connection on a principal
$G$-bundle $P$ over a manifold $\Sigma$ which represents `space', and
the field ${\tilde E}$ is the canonically conjugate momentum (the
tilde means that it is a densitized vector field).  For technical
reasons we assume that $\Sigma$ is real-analytic and $G$ is a compact
connected Lie group.  Actually the assumption that $\Sigma$ is
analytic is unnecessary \cite{BS}; much of what we do in this paper
generalizes to the smooth case, but analyticity makes things a bit
simpler.

The classical configuration space of our theory is the the space of
smooth connections on $P$, which we denote as $\cal A$.  In the case
of canonical quantization of a theory with a finite number of degrees
of freedom one normally represents quantum states by square-integrable
functions on the classical configuration space. The Hilbert space of
states is then constructed as the space of such functions with the
inner product being defined by the usual integral over the
configuration space, $\IP{\psi}{\phi} = \int \overline{\psi} \phi\,
d\mu$, where $d\mu$ is some measure on the configuration
space. However, in the case of theories with an infinite number of
degrees of freedom, which is of interest for us here, there often does
not exist an appropriate measure on the classical configuration space.
For example, there is no `Lebesgue measure' on an infinite-dimensional
vector space.  However, one can often complete the classical
configuration space in some topology and construct a suitable measure
on the resulting `quantum configuration space'.  In field theories,
this larger space usually includes distributional fields in addition
to smooth fields.  Also, it is common in field theories for the
classical configuration space to have zero measure with respect to the
measure on the quantum configuration space.

The quantum configuration space depends on a choice of a 
functions on the classical configuration space, which play
the role of distinguished observables.   In our case we work with 
(smooth) `cylinder functions' on 
$\cal A$, that is, functions dependin smoothly on the holonomy of the
connection $A$ along finitely many analytic paths.
In other words, $\Psi$ is a cylinder function if it is of
the form 
\[   \Psi(A) = \psi({\cal P} \exp \int_{e_1} A, \dots,
 {\cal P} \exp \int_{e_n} A) \]
for some analytic paths 
$e_1,\dots,e_n$ in $\Sigma$ and some smooth function $\psi$ on $G^n$.
The most well-known functions of this form are the Wilson loops, introduced
as observables for quantum gravity by Rovelli and Smolin \cite{Loops}
in their original paper on the loop representation.  Unlike the
Wilson loops, the above cylinder functions are not necessarily 
gauge-invariant.  This allows them to serve as a complete set of
functions on $\cal A$.  

One can complete the algebra of cylinder functions with respect to 
the $\sup$ norm to obtain a commutative C*-algebra $\Fun({\cal A})$ for which
the $\ast$-operation is just pointwise complex conjugation.  
The Gelfand-Naimark theorem then tells us that this C*-algebra is 
isomorphic to the algebra of all continuous functions on its spectrum.
We take the spectrum of $\Fun({\cal A})$ to be the quantum configuration
space of our theory.  We denote this space by $\a$, because it contains
the classical configuration space $\cal A$ as a dense subset.  

Elements of the quantum configuration space are called `generalized
connections', and can be described as follows.   First,
define a `transporter' from the point $p$ to the
point $q$ to be a map from $P_p$ to $P_q$ that commutes with 
the right action of $G$ on the bundle $P$.  
If we trivialize the bundle over $p$ and $q$, we can think of
such a transporter simply as an element of $G$.
A `generalized connection' $A$ is a map assigning to 
each oriented analytic path $e$ in $\Sigma$ a parallel transporter
$A_e \maps P_p \to P_q$, where $p$ is the initial point of the
path $e$ and $q$ is the final point.   We require that $A$ satisfy
certain obvious consistency conditions: $A$ should assign the
same transporter to two paths that differ only by an orientation-preserving
reparametrization, it should assign to the inverse of any path the 
inverse transporter, and it should assign to the composite of 
two paths the composite transporter.  

An ordinary smooth connection $A$ gives a generalized connection where
the parallel transporter $A_e$ along any path $e$ is simply the holonomy of
$A$ along this path, so $\cal A \subseteq \a$.  
Also, any cylinder function extends to a continuous 
function on $\a$ by setting
\[      \Psi(A) = \psi(A_{e_1},\dots, A_{e_n}) \]
for any generalized connection $A$.   This lets us think of cylinder
functions as functions on the quantum configuration space.

There is a natural measure $\mu_0$ on $\a$, which comes from 
Haar measure on the group $G$.   One can show that any
reasonable measure on $\a$ is determined by the values of
the integrals of all cylinder functions.  
Thus, to define the measure $\mu_0$ we specify the values of the
integrals $\int \Psi \,d\mu_0$ for all cylinder functions $\Psi$ given as
above.  Such an integral is 
defined as the integral over the corresponding copies of the group $G$:
\[  \int \Psi\, d\mu_0 = 
\int \psi(g_1,\dots, g_n) \, dg_1 \cdots dg_n \]
where $dg$ denotes the normalized Haar measure on $G$.  
The measure $\mu_0$ has three
important properties.   First, it is gauge-invariant.
More precisely, note that any gauge transformation $g$ acts
on $A \in \a$ to give a generalized connection $A'$ with
$A'_e = g(q)^{-1}\circ A_e \circ g(p)$, 
where $p,q$ are the final and the initial point of the 
path $e$, respectively.   This gives an action of the group
$\g$ of gauge transformations of $P$ on the space $\a$, and
this action preserves the measure $\mu_0$.   Second, $\mu_0$
is diffeomorphism-invariant.  More precisely, $\mu_0$ is invariant under all
automorphisms of the bundle $P$, not necessarily acting as the identity 
on the base space $\Sigma$.  Third, $\mu_0$ is
strictly positive, meaning that $\int \Psi\, d\mu_0 > 0$ for
any nonnegative integrable function on $\a$ except the function $0$. 

We define the kinematical Hilbert space $L^2(\a)$ to be the space of
functions on $\a$ that are square-integrable with respect to the
measure $\dm0$.  This space is not the physical state space, since it
contains states that are not invariant under the `gauge' symmetries of
our theory, that is, gauge transformations and diffeomorphisms of
spacetime.  One way to try to find physical states is to look for
solutions of quantum constraints in $L^2(\a)$.  In general, the
solutions may not live in $L^2(\a)$, but in some completion thereof,
but for the Gauss law constraint this problem does not occur: there is
a large subspace of $L^2(\a)$, the `gauge-invariant Hilbert space',
consisting of gauge-invariant square-integrable functions on $\a$.
Alternatively, since the measure $\mu_0$ is gauge-invariant, it gives
rise to a well-defined measure on the space $\ag$ of generalized
connections modulo gauge transformations.  Then we may equivalently
define the gauge-invariant Hilbert space to be the space $L^2(\ag)$ of
square-integrable functions on $\ag$.

We can construct an explicit basis of $L^2(\ag)$ using `spin networks'
\cite{B,RS}.  A spin network is a triple $(\gamma,\rho,\iota)$
consisting of:
\vskip 1em
\noindent (1) a graph $\gamma$ analytically embedded in $\Sigma$,

\noindent (2) a labelling $\rho$ of each edge $e$ of $\gamma$ with an
irreducible representation $\rho_e$ of $G$,

\noindent (3) a labelling $\iota$ of each
vertex $v$ of $\gamma$ with an intertwining operator $\iota_v$.
\vskip 1em
\noindent Here the edges of the graph $\gamma$ are assumed to be 
unparametrized but oriented,
and $\iota_v$ is an interwining operator from the
tensor product of the representations corresponding to the incoming
edges at the vertex $v$ to the tensor product of the representations
labeling the outgoing edges.  (For more details see references \cite{AL,B}.
Here we allow graphs with isolated vertices, i.e., vertices that are not 
vertices of any edge.  This is unimportant now but will be important 
when we come to fermions.) 
Without loss of generality, spin networks $\Gamma$ are always 
assumed to satisfy a fourth non-degeneracy condition:
\vskip 1em
\noindent (4) All representations $\rho_e$ are
nontrivial and $\gamma$ is a `minimal' graph, in the sense that it
connot be obtained from another graph $\gamma'$ by subdividing edges
of $\gamma'$.
\vskip 1em

The `spin network state' $\Psi_\Gamma$ is a gauge-invariant cylinder
function on $\a$ constructed from the spin network $\Gamma$ as follows:
\[
\Psi_\Gamma(A) = \bigl [\bigotimes_e\,
\rho_e(A_e)\bigr ]\cdot\bigl [\bigotimes_v\,\iota_v \bigr ],
\]
where `$\cdot$' stands for contracting, at each vertex $v$ of
$\gamma$, the upper indices of the matrices corresponding to the
incoming edges, the lower indices of the matrices assigned to the
outgoing edges, and the corresponding indices of the intertwining operator
$\iota_v$.  Being gauge invariant, such states lie in $L^2(\ag)$.  We
obtain an orthonormal basis of $L^2(\ag)$ if we use spin network
states corresponding to all possible choices of $\gamma$ and $\rho$ and a
choice of an orthonormal basis of intertwining operators $\iota_v$ for each
vertex $v$ and each choice of representations labeling incident edges.

The states in $L^2(\ag)$ are still not invariant under the action of 
diffeomorphisms of $\Sigma$.  There is a natural unitary representation of
Diff($\Sigma$) on $L^2(\ag)$. For instance, the action of the 
operator $U(\phi)$ corresponding to a diffeomorphism $\phi$ of
$\Sigma$ on any spin network state is given by
\[
U(\phi)\Psi_{(\gamma,{\rho},{\iota})} = 
\Psi_{(\phi\gamma,\phi{\rho},\phi{\iota})}.
\]
Here $\phi\gamma$ is the image of the graph $\gamma$ under the
diffeomorphism $\phi$, and $\phi\rho,\phi\iota$ are the corresponding
representations and intertwining operators associated with the new
graph $\phi\gamma$.

To find diffeomorphism-invariant states one has to impose the quantum
diffeomorphism constraint.  There are very few
diffeomorphism-invariant states in $L^2(\ag)$, so here we have to
follow the strategy of looking for solutions in some larger space.
One often needs to do this when solving quantized constraint
equations, and the larger space is usually chosen to be a space of
linear functionals on some dense subspace of the initial Hilbert
space. That is, we should choose some dense subspace $\cal C$ of the
gauge-invariant Hilbert space $L^2(\ag)$, equip it with some topology
in which it is complete, and look for solutions in the topological
dual ${\cal C}^*$.  Note that ${\cal C} \subset L^2(\ag) \subset {\cal
C}^*$.

Since there is a simple geometrical action of diffeomorphisms on
cylinder functions, it is natural to choose the subspace of
gauge-invariant cylinder functions as $\cal C$.  This has a natural
topology in which it is complete, the inductive limit topology.  To
solve the diffeomorphism constraint, we then seek
diffeomorphism-invariant vectors in $\cal C^*$.

We obtain these by averaging gauge-invariant cylinder functions over
the action of the diffeomorphism group, following the
procedure of Ashtekar, Lewan-\break dowski, Marolf, Mour\~ao and Thiemann
\cite{ALMMT}.  It is easiest to average a
special sort of gauge-invariant cylinder function, namely a spin
network state $\Psi_\Gamma$, where $\Gamma = (\gamma,\rho,\iota)$ is
some spin network.  For technical reasons we assume $\gamma$ is 
`type I', i.e., that for every edge of $\gamma$ there is an
analytic function on $\Sigma$ that vanishes only on the maximal
analytic curve extending that edge.  
Let $S(\gamma)$ be any set of diffeomorphisms of
$\Sigma$ with the following property: for any graph $\gamma'$ which
equals $\phi \gamma$ for some $\phi \in {\rm Diff}(\Sigma)$, there is
a unique diffeomorphism $\phi \in S(\gamma)$ with $\gamma' = \phi
\gamma$.  Also let $GS(\gamma)$ be the group of `graph symmetries' of
$\gamma$, that is, the group ${\rm Iso}(\gamma)/{\rm TA}(\gamma)$,
where ${\rm Iso}(\gamma)$ is the group of diffeomorphisms mapping
$\gamma$ to itself, and ${\rm TA}(\gamma)$ is the subgroup fixing
each edge of $\gamma$.  
We may define an element $\overline{\Psi}_\Gamma \in {\cal C}^*$
by: 
\[\overline{\Psi}_\Gamma(\Phi) =
\sum_{\phi_1\in S(\gamma)} \sum_{[\phi_2]\in GS(\gamma)} 
\IP{\Psi_{(\phi_1 \circ \phi_2)\Gamma}}{\Phi},\]
where $\Phi \in \cal C$, and where we choose one representative
$\phi_2$ for each equivalence class $[\phi_2] \in GS(\gamma)$.
It is easy to check that $\overline{\Psi}_\Gamma$
is diffeomorphism-invariant.  
More generally, suppose the cylinder function $\Psi$ is a
(possibly infinite) linear combination of such spin network states 
$\sum_\Gamma a_\Gamma \Psi_\Gamma$.  Then we define
$\overline{\Psi} \in \cal C^*$ by:
\[ \overline{\Psi}(\Phi) = 
\sum_\Gamma a_\Gamma \overline{\Psi}_\Gamma(\Phi) .\]
It is easy to see that the sum converges and defines a
diffeomorphism-invariant element $\overline \Psi \in \cal C^*$.

We may define the inner product of diffeomorphism-invariant
vectors of this form by 
\[  \IP{\overline{\Psi}}{\overline{\Phi}} = 
\overline{\Psi}(\Phi). \]
Completing this space of such vectors in this inner product we thus
obtain a Hilbert space, the `diffeomorphism-invariant Hilbert space'
$\hdiff$.   The diffeo-\break morphism-invariant spin network states $\overline
\Psi_\Gamma$ form an orthogonal (but not orthonormal) basis of 
$\hdiff$ as we let $\Gamma$ range over a set of spin networks containing
one from each diffeomorphism equivalence class (but restricting
ourselves to those whose graph is type I).  For any spin
network $\Gamma$, the diffeomorphism-invariant state
$\overline{\Psi}_\Gamma$ takes a zero value on spin networks that do
not belong to the same equivalence class, and a nonzero value on spin
networks from the same equivalence class.

In short, the main steps of the quantization procedure include: (i)
construction of the kinematical Hilbert space of states $L^2(\a)$,
which requires the specification of a quantum configuration space
together with an appropriate measure on this space; (ii) construction
of the gauge-invariant Hilbert space $L^2(\ag)$ (and, conveniently, an
explicit basis for this space), and (iii) construction of the
diffeomorphism-invariant Hilbert space $\hdiff$.  In what follows
we generalize all these steps to the case of theories involving fermions.
Finally, there is the problem of imposing the Hamiltonian constraint 
and the problem of finding solutions to this constraint.  
In this paper we do not enter into this all-important problem.

In the next section we describe the class of theories which are
considered in this paper and give several examples of physically
interesting theories belonging to this class.  Their quantization will
be discussed in the following sections.

\section{The class of theories}
\label{sec:3}

In this paper we consider a special class of theories with fermionic
degrees of freedom. Each theory belonging to this class is specified,
first, by: (i) a principal $G$-bundle $P \to \Sigma$, where the gauge
group $G$ is a compact connected Lie group and $\Sigma$ is a
real-analytic manifold.   We also need to specify (ii) a finite list $I$
of irreducible continuous unitary representations of $G$.  Each
representation in $I$ corresponds to an elementary fermion 
appearing in the theory.  Using any representation $\rho \in I$ we 
may associate to $P$ a vector bundle $P \times_G \rho$ over $\Sigma$.   
It will be convenient to lump these together by forming a single vector bundle
\[         F = P \times_G \bigl( \bigoplus_{\rho \in I} \rho\bigr) \,,  \]
sections of which simultaneously describe all the fermionic fields 
in the theory.

>From this data we can construct the classical configuration space and
phase space of the theory.  We have already described the classical
configuration space for the gauge fields; this is the space $\a$ of
smooth connections on $P$.  The corresponding classical phase space is
the space of pairs $(A,\tilde E)$ where $A$ is a smooth connection on
$P$ and $E$ is a smooth $\Ad P$-valued vector field of density weight
1.  We denote this classical phase space as $T^\ast {\cal A}$, since we
can think of $\tilde E$ as a cotangent vector to $\cal A$ satisfying
certain smoothness conditions.  Both $\cal A$ and $T^\ast \cal A$
become infinite-dimensional smooth manifolds in a natural way.

Similarly, the classical configuration space for the fermionic fields
is the space $\cal F$ of smooth sections of $F$, and the classical
phase space for the fermionic fields is space of pairs consisting of a
smooth section of $F$ together with a densitized smooth section of
$F^\ast$ with density weight 1.  We denote this classical phase space
as $T^\ast \cal F$.  Both $\cal F$ and $T^\ast \cal F$ become
infinite-dimensional topological vector spaces in a natural way.

The classical configuration space for the whole theory is thus the
product ${\cal A} \times {\cal F}$, while the classical phase space is
$T^\ast {\cal A} \times T^\ast \cal F$.  In down-to-earth terms, a
point in the classical phase space is simply a list of pairs
$(A,\tilde{E}), (\xi,{\tilde \pi}), \dots, (\eta,{\tilde\omega})$.
Here $A$ is the connection on $P$ and $\tilde{E}$ is its canonically
conjugate momentum.  Similarly, the fields $\xi, \dots, \eta$ are
fermionic configuration fields corresponding to the representations in
$I$, while the fields ${\tilde\pi}, \dots, {\tilde\omega}$ are their
canonically conjugate momenta.

In what follows we shall treat the fermionic fields as
Grassmann-valued, in order to guarantee that the Pauli principle holds
for our fermions.  This amounts to treating $\cal F$ as a
supermanifold with all odd coordinates \cite{DeWitt}.  Thus we define
the algebra $C^\infty({\cal F})$ of `smooth functions' on $\cal F$ to
be the exterior algebra $\Lambda {\cal F}^\ast$ generated by the
continuous linear functionals on $\cal F$.  Note that these are not
functions on $\cal F$ in the standard sense, but only in the sense of
supermanifold theory.  Similarly, we define the algebra
$C^\infty(T^\ast {\cal F})$ of `smooth functions' on the fermionic
classical phase space $T^\ast {\cal F}$ to be the exterior algebra
generated by the continuous linear functions on $T^\ast {\cal F}$.
We also define `smooth functions'
on the configuration space and phase space of the whole theory as
follows:
\[      C^\infty({\cal A} \times {\cal F}) = 
C^\infty({\cal A}) \tensor C^\infty({\cal F}), \]
\[     C^\infty(T^\ast {\cal A} \times T^\ast {\cal F}) = C^\infty(T^\ast 
{\cal A}) \tensor C^\infty(T^\ast {\cal F}) .\] 
The latter algebra is the algebra of classical observables of the theory.

Theories may have the same phase space and differ only in the form of
the Hamiltonian, so a theory is finally determined by: (iii) the
Hamiltonian.  Since we shall not actually treat dynamics in this
paper, we shall not be very precise about the class of allowed
Hamiltonians, but we are interested in those for which 
the action can be written as follows:
\begin{equation}
S = \int dt \int d^3x \bigl ( \mathop{\rm Tr}\,\tilde{E}^a{\cal L}_t\A + 
{\tilde \pi}{\cal L}_t\xi + {\tilde \omega}{\cal L}_t\eta + \cdots +
\ut{N}\,{\tilde{\tilde H}} + N^a\,{\tilde H}_a + 
\mathop{\rm Tr}\,{\bf N}\,{\tilde G}.
\nonumber
\end{equation}
Here ${\cal L}_t$ stands for the time derivative and the quantities
$\ut{N},N^a,$ and ${\bf N}$ are Lagrange multipliers. The quantities
${\tilde{\tilde H}}, {\tilde H_a},$ and ${\tilde G}$ are functionals
on the phase space corresponding to the Hamiltonian, diffeomorphism
and Gauss law constraints.  Here tildes over the momentum fields keep
track of density weights: a single tilde over a symbol stands for a
density of weight $1$, a double tilde over a symbol stands for a
density of weight $2$, and a single tilde under a symbol denotes a
density of weight $-1$.

Once the canonical variables are chosen (or in other words, once the
phase space is chosen), a theory is determined solely by the
Hamiltonian.  The other two constraint functionals are determined by
the requirement that they generate diffeomorphisms and gauge
transformations on the phase space of the theory.

Let us give two examples.

\paragraph{Example A.}

An example of such a theory was considered in \cite{RovFerm}. It
describes a massless fermionic field interacting with a gravitational
field in four-dimensional (Riemannian) spacetime, with the
gravitational field being described using a chiral spin connection.
The gauge group is $\SU(2)$, and the gravitational degees of freedom
are described by the $\SU(2)$ connection $A_a$ and its canonically
conjugate momentum field $\tilde{E}^a$.  The only fermionic field of
the theory, $\xi$, takes values in the spin-$1/2$ representation.  The
Hamiltonian constraint of the theory consists of two parts:
\[
\tilde{\tilde H} = 
{1\over 2}\,\mathop{\rm Tr}(\tilde{E}^a\tilde{E}^b{F}_{ab}) +
\tilde{E}_A^{a\;B}\,{\cal D}_a\xi_B{\tilde \pi}^A,
\]
where the first part is the Hamiltonian of the gauge field and the second
part is responsible for the dynamics of the fermionic field. 
Here $F_{ab}$ is the curvature of the connection field $A_a$, and 
${\cal D}_a = \partial_a+A_a$ is the covariant
derivative operator. This theory
is the simplest theory of gravity coupled to matter. 

\paragraph{Example B.}

Another example is the theory describing massive fermions interacting
both gravitationally and electromagnetically, which was considered in
\cite{EMF}. The gauge group here is $\SU(2)\times \U(1)$.  The
gravitational and electromagnetic degrees of freedom are described 
together by a $\SU(2) \times \U(1)$ connection field $A_a$ and 
canonically conjugate momentum field $\tilde{E}^a$.
The fermionic degrees of freedom are described 
by two Grassmann-valued fields $\xi,\eta$ taking values in the
representations of spin $1/2$ and charge $\pm 1$.
The Hamiltonian of the
theory consists of the following three parts.   The part generating the 
dynamics of the gravitational field is
\[
{\tilde{\tilde H}}{}_{\rm grav} = {1\over2}\ut{\eta}_{\,abc}
\mathop{\rm Tr}(\E^a\E^b\tilde{B}^{c}),
\]
where the magnetic field $\tilde{B}^{a}$ is the dual of
the curvature 
$F_{ab} = \ut{\eta}_{\,abc}\tilde{B}^{c}$, 
$\ut{\eta}_{\,abc}$ being the totally antisymmetric tensor of 
weight $-1$.  The part generating the dynamics of the electromagnetic
field is
\[
{\tilde{\tilde H}}{}_{\rm em} = 
\]
\[
{1\over 32} \sigma^{-2}
\ut{\eta}_{\,abe}
\ut{\eta}_{\,cdf}\,\mathop{\rm Tr}(\E^{a}\E^{c})
\Bigl [\mathop{\rm Tr}(\E^{b}\E^{d})
      \mathop{\rm Tr}(\tilde{B}^{e}) \mathop{\rm Tr}(\tilde{B}^{f}) 
   -\mathop{\rm Tr}(\E^{b}\E^{d})
      \mathop{\rm Tr}(\E^{e}) \mathop{\rm Tr}(\E^{f})  \]
\[
   -\mathop{\rm Tr}(\E^{b}) \mathop{\rm Tr}(\E^{d})
      \mathop{\rm Tr}(\tilde{B}^{e}) \mathop{\rm Tr}(\tilde{B}^{f}) 
\Bigr ]
\]
where $\sigma$ is the square root of the determinant of the
metric, as defined using the $\tilde E$ field.
Finally, the part of the Hamiltonian responsible for the fermionic fields is
\[
{\tilde{\tilde H}}{}_{\rm ferm} = \bigl [\E^{a\;B}_{\,A} - 
{1\over2}\mathop{\rm Tr}(\E^a)\delta_{A}^{\,B}\bigr ]
\bigl ({\cal D}_{a}\xi_B{\tilde \pi}^A +
{\cal D}_{a}\eta^A{\tilde \omega}_B\bigr ) +
im\,\bigl [\tilde{\pi}^{A}\tilde{\omega}_{A} -
(\sigma)^{2}\eta^A\xi_A \bigr ]
\]
where $m$ is the fermion mass.
\vskip 1em

In both the theories above particles appear in the theory with their
antiparticles.  Mathematically this is manifested by the fact that
each irreducible representation that appears in the theory appears
with its dual.  However, this comes about in very different ways for
the two different theories above.  The $\SU(2)$ theory in Example A
contains only one fermionic field, but this field transforms under a
representation that is isomorphic to its dual.  It describes a
particle that is its own antiparticle.  In Example B we have two
different fermionic fields transforming under dual representations of
the group $\SU(2) \times \U(1)$.  This theory describes a particle
that is distinct from its antiparticle.

One can see that representations must appear with their duals
in order to construct a gauge-invariant mass term
in the Hamiltonian, or any other gauge-invariant expression bilinear
in the configuration fermionic fields.  It is sensible therefore to
limit our attention to theories for which each irreducible
representation of $G$ appears in the list $I$ together with its dual.
Actually, to streamline the exposition, we shall consider only the
case where particles are distinct from their antiparticles.  That is,
we assume: (iia) the list $I$ of representations of $G$ is of the form
$(\rho_1,\rho_1^\ast, \dots ,\rho_n, \rho_n^\ast)$.  It is not
difficult to extend our analysis to the case of particles that are
their own antiparticles, and we mention a few of the changes that need
to be made as we come to them.

Having described the class of theories which are the subject of this
paper we are now ready to proceed with the quantization program.  For
this we need to construct fermionic path observables, which are the
basic building blocks of the quantum theory.

\section{Fermionic path observables}
\label{sec:4}

We now introduce `path observables', which play in our
theories a role similar to that of the standard Wilson loop
observables in theories of connections.  We start with by
describing some path observables that are functions on the
classical configuration space ${\cal A} \times {\cal F}$.  These observables
are built from the holonomy of the connection along a path
in space together with fermion fields at the endpoints of this path.

Note first that given two fermionic fields $\xi,\eta$ transforming under dual
representations $\rho$ and $\rho^*$ of the gauge group, one can construct a
gauge-invariant quantity from these fields using the
$G$-invariant bilinear pairing between $\rho^\ast$ and $\rho$.  
This quantity depends on the values $\xi_p ,\eta_p$ of these fields
at a point $p\in\Sigma$ and is given simply by $(\xi_p,\eta_p)$.
More generally, given points $p$ and $p'$ and a path $e$ from
$p$ to $p'$, one can parallel transport $\xi_p$ to $p'$ along the
path $e$ using the connection $A$ and then pair it with $\eta_{p'}$, 
obtaining the gauge-invariant quantity 
\[   (\xi_p|e|\eta_{p'}) = (\rho({\cal P}\exp{\int_e A})\xi_p,\,\eta_{p'}) \]
which we call a `configurational path observable'.   It is also
helpful to have a graphical notation for a configurational path
observable, in which we represent the oriented path $e$ by a line
with an arrow on it, and draw the fermionic fields $\xi$ and $\gamma$ 
as dots at the endpoints:
\begin{equation}
(\xi_p|e|\eta_{p'}) \qquad\qquad\qquad {e\atop\kd{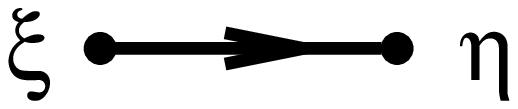}}
\label{lvar1}
\end{equation}
We may think of the dot labelled by $\xi$ as a particle and the
the dot labelled by $\eta$ as its antiparticle.  Configurational path
observables can be introduced for any representation in the list $I$ together
with its dual, or in other words, for any particle in the the theory
together with its antiparticle.  

Note that the configurational path observables are even since they
involve a product of two fermion fields.  As a result they commute.
We may also think of them as functions on the classical phase space,
but since they involve no momenta they all Poisson-commute.  Let us
now introduce the basic path observables involving fermionic momentum
fields and describe the Poisson algebra these observables generate.

Recall that the momentum field $\tilde\pi$ canonically conjugate to 
$\xi$ transforms under the dual representation.  This means that one can 
apply the above construction to the $\xi,\tilde{\pi}$ fields. 
Namely, given two points $p,p'$ and a path $e$ connecting them,
one can construct the quantity 
\[ (\xi_p|e|\tilde{\pi}_{p'})= 
(\rho[{\cal P}\exp{\int_e A}]\xi_p,\, \tilde{\pi}_{p'}), \]  
which is gauge invariant and depends only on the values of fields
$\xi,\tilde {\pi}$ at the endpoints $p,p'$ of the path $e$ and on
the holonomy of the connection field along $e$.

The quantity $(\xi|\gamma|\tilde{\pi})$ is almost the one which we
need. However, the Poisson bracket of this quantity with a
configurational path observable $(\xi|\gamma|\eta)$ is a distribution.
Following \cite{EMF}, let us introduce some `averaged' momentum
observables in such a way that the Poisson bracket of a
configurational path observable with a momentum observable is again a
configurational path observable.  In other words, let us introduce
momentum observables in such a way that the resulting Poisson algebra
contains no distributions.  To do this we choose an arbitrary rule
which specifies a path $e_{pp'}$ from $p$ to $p'$ given points
$p,p'$ in some region ${\cal R} \subset \Sigma$, an open set 
whose closure is compact.  Having this rule at our disposal, we can integrate
$(\xi|e_{pp'}|\tilde{\pi})$ over ${\cal R}$ as a function of $p'$.  
We obtain the quantity
\begin{equation}
\int_{\cal R} d^np'\, (\xi_p|e_{pp'}|\tilde{\pi}_{p'}) \qquad\qquad\qquad 
{e\atop\kd{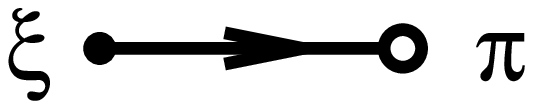}}
\label{lvar2}
\end{equation}
Here in our graphical notation we use a dot as before to stand for the
fermionic configuration field, but use a little circle to
represent the averaged fermionic momentum field.  The result is a
gauge-invariant quantity, which depends on values of the momentum
field $\tilde{\pi}$ and the connection field $A$ over a region of the
spatial manifold, on the value of $\xi$ at the point $p$, and on a
rule $(p,p')\mapsto e_{pp'}$.

Similarly, one can introduce a quantity
\begin{equation}
\int_{\cal R} d^np' \,(\tilde{\omega}_{p'}|e_{p'p}|\eta_p)
\qquad\qquad\qquad 
{e\atop\kd{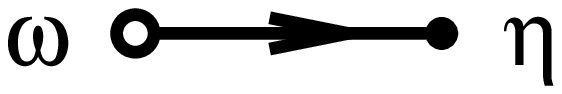}}
\label{lvar3}
\end{equation}
depending on the momentum field $\tilde{\omega}$ canonically 
conjugate to $\eta$ (or, physically, the momentum field of the antiparticle). 
One can also construct quantities involving two momentum fields
\begin{equation}
\int_{\cal R} d^np \int_{\cal R'} 
d^np'(\tilde{\omega}_p|e_{pp'}|\tilde{\pi}_{p'}), 
\qquad\qquad\qquad
{e\atop\kd{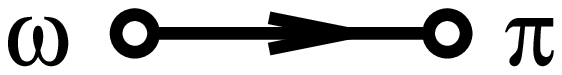}}
\label{lvar4}
\end{equation}
Again we represent averaged momentum fields by little circles. 

Note that were we to consider a theory in which $\xi$ coincided with
its antiparticle $\eta$, there would be only one path observable 
linear in the momentum field, instead of two different quantities 
(\ref{lvar2}) and (\ref{lvar3}).  Some of the Poisson brackets below
would be different in this case, but the necessary modifications
are not difficult.

The quantities we have described so far are not invariant under the
Hermitian conjugation operation `$\dagger$'.  This operation is
defined when the gauge fields of our theory include gravity, and the
crucial property it satisfies is given by equation (\ref{dagger}).
Let us now introduce a set of `real' observables, by which we mean
functions on the classical phase space that are preserved by this
$\dagger$ operation.  Since Hermitian conjugation sends fermionic
momentum fields into configuration fields, real observables should
involve both fields.  Let us introduce the following observables which
are linear in the fermionic momenta:
\begin{eqnarray}
\int_{\cal R} d^np\, (\xi_p, \tilde{\pi}_p), \qquad\qquad\qquad 
{\kd{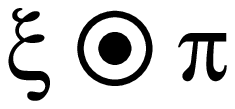}}
\label{lvar5} \\
\int_{\cal R} d^np \,(\tilde{\omega}_p, \eta_p), \qquad\qquad\qquad 
{\kd{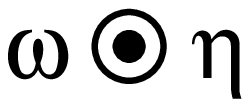}}
\label{lvar6}
\end{eqnarray}
where $(\cdot,\cdot)$ is the bilinear pairing between the
dual representations $\rho$ and $\rho^*$. 
As we shall see, when quantized these observables play the role of number
operators measuring the number of particles of the
appropriate sort in the region $\cal R$.  The graphical representation 
that we use for these observables is as shown above.

The quantities we have introduced form a set of functions on the phase
space of the theory.  Let us now describe the Poisson algebra which
these quantities generate (we also give a graphical
description). First, the Poisson brackets of the observables linear in
the momentum (\ref{lvar2}-\ref{lvar3}) with the configurational path
observables (\ref{lvar1}) are again configurational path observables.
The following identities can be verified \cite{EMF}:
\begin{eqnarray}
\left \{\;\int_{\cal R} d^np' \, (\xi_p|e_{pp'}|\tilde{\pi}_{p'})\;,\;
(\xi_q|f|\eta_{q'})\;\right \} &=& (\xi_p|e_{pq}\circ f|\eta_{q'}),
\nonumber \\
\left \{ {e\atop\kd{lvar2}}, 
{f\atop\kd{lvar1}} \right \} &=& {e\circ f\atop\kd{lvar1}}
\label{br1}
\end{eqnarray}
\begin{eqnarray}
\left \{\;\int_{\cal R} d^np' (\tilde{\omega}_{p'}|e_{p'p}|\eta_p)\;,\;
(\xi_q|f|\eta_{q'})\;\right \} &=& (\xi_q|f\circ e_{q'p}|\eta_{p}).
\nonumber \\
\left \{ {e\atop\kd{lvar3}},{f\atop\kd{lvar1}} \right \} &=& 
{f\circ e\atop\kd{lvar1}}
\label{br2}
\end{eqnarray}
Here we use `$\circ$' to denote composition of paths.  Also, it is
assumed in (\ref{br1}) that $q\in{\cal R}$.  If the point $q$ lies
outside of the region $\cal R$ the Poisson bracket is zero.
Similarly, in (\ref{br2}) we assume $q'\in {\cal R}$; otherwise the
Poisson bracket is zero.  Note that if the result is not zero, in both
cases it is given by a configurational path observable with the path
being the composite of the paths $e, f$ compatible with their
orientation.  

The Poisson bracket of a momentum observable (\ref{lvar2})
with another observable of this sort gives
\begin{eqnarray}
\left \{\;\int_{\cal R} d^np' \,(\xi_p|e_{pp'}|\tilde{\pi}_{p'})\;,\;
\int_{\cal R'} d^nq'\, (\xi_q|f_{qq'}|\tilde{\pi}_{q'})\;\right \} = 
\nonumber \\
\int_{\cal R'} 
d^nq' \, (\xi_p|e_{pp'}\circ f_{p'q'}|\tilde{\pi}_{q'})\;+\;
\int_{\cal R} d^np'\, (\xi_q|f_{qq'}\circ e_{q'p'}|\tilde{\pi}_{p'}).
\label{br3} \\
\left \{ {e\atop\kd{lvar2}}, {f\atop\kd{lvar2}} \right \} = 
{e\circ f\atop\kd{lvar2}}+{f\circ e\atop\kd{lvar2}}
\nonumber
\end{eqnarray}
Similarly, for (\ref{lvar3}) we get
\[
\left \{\;\int_{\cal R} d^np' \,
(\tilde{\omega}_{p'}|e_{p'p}|\eta_p)\;,\;
\int_{\cal R'} d^nq'\, (\tilde{\omega}_{q'}|f_{q'q}|\eta_q)\;\right \} = \]
\begin{eqnarray}
\int_{\cal R}
d^np'\, (\tilde{\omega}_{p'}|e_{p'p}\circ f_{pq}|\eta_q)\;+\;
\int_{\cal R'} 
d^nq'\, (\tilde{\omega}_{q'}|f_{q'q}\circ e_{qp}|\eta_p).
\label{br4} \\
\left \{ {e\atop\kd{lvar3}}, {f\atop\kd{lvar3}} \right \} = 
{e\circ f\atop\kd{lvar3}}+{f \circ e\atop\kd{lvar3}}
\nonumber
\end{eqnarray}
where we assume that the point where
the configuration field in each of the observables is evaluated lies inside
the region over which the momentum field in the other is smeared; otherwise
one or both terms of the Poisson bracket vanishes.

The Poisson bracket of (\ref{lvar4}) and a configuration
observable also consists of two terms:
\begin{eqnarray}
\left \{\;
\int_{\cal R} d^np 
\int_{\cal R'} d^np'(\tilde{\omega}_p|e_{pp'}|\tilde{\pi}_{p'})\;,\;
(\xi_q|f|\eta_{q'})\;\right \} = \nonumber \\
\int_{\cal R} d^np (\tilde{\omega}_{p}|e_{pq}\circ f|\eta_{q'}) +
\int_{\cal R'} d^np' (\xi_q|f\circ e_{q'p'}|\tilde{\pi}_{p'}).
\label{br5} \\
\left \{ {e \atop\kd{lvar4}}, {f\atop\kd{lvar1}} \right \} = 
{e \circ f\atop\kd{lvar3}}+{f \circ e\atop\kd{lvar2}}
\nonumber
\end{eqnarray}
where again we assume $q\in{\cal R'}$ and $q'\in{\cal R}$.
For the Poisson brackets between (\ref{lvar4}) and observables linear
in momentum we have
\begin{eqnarray}
\left \{\;
\int_{\cal R} d^np 
\int_{\cal R'} d^np'(\tilde{\omega}_p|e_{pp'}|\tilde{\pi}_{p'})\;,\;
\int_{\cal R''} d^nq' (\xi_q|f_{qq'}|\tilde{\pi}_{q'})\;\right \} = 
\nonumber \\
\int_{\cal R} d^np 
\int_{\cal R''} d^nq'
(\tilde{\omega}_p|e_{pq}\circ f_{qq'}|\tilde{\pi}_{q'}),
\label{br6} \\
\left \{ {e\atop\kd{lvar4}}, {f\atop\kd{lvar2}} \right \} = 
{e\circ f\atop\kd{lvar4}}
\nonumber
\end{eqnarray}
\begin{eqnarray}
\left \{\;
\int_{\cal R} d^np 
\int_{\cal R'} d^np'(\tilde{\omega}_p|e_{pp'}|\tilde{\pi}_{p'})\;,\;
\int_{\cal R''} d^nq' (\tilde{\omega}_{q'}|f_{q'q}|\eta_{q})\;\right \} 
= \nonumber \\
\int_{\cal R''} d^nq' 
\int_{\cal R'} d^np'
(\tilde{\omega}_{q'}|f_{q'q}\circ e_{qp'}|\tilde{\pi}_{p'}).
\label{br7}\\
\left \{ {e\atop\kd{lvar4}}, {f\atop\kd{lvar3}} \right \} = 
{f\circ e\atop\kd{lvar4}}
\nonumber
\end{eqnarray}
where in (\ref{br6}) we assume $q'\in{\cal R'}$, and in
(\ref{br7}) we assume $q\in{\cal R}$; otherwise the brackets vanish.

One can also write down Poisson brackets between 
(\ref{lvar1}-\ref{lvar4}) and the real quantities (\ref{lvar5}-\ref{lvar6}).
For example, the Poisson brackets between configurational path
observables and these real quantities are given by
\begin{eqnarray}
\left \{\;\int_{\cal R} d^np (\xi_p, \tilde{\pi}_{p})\;,\;
(\xi_q|e|\eta_{q'})\;\right \} &=& (\xi_q|e|\eta_{q'}),
\nonumber \\
\left \{ {\pic{lvar5}{13}{-10}{2}{2}}, 
{e\atop\kd{lvar1}} \right \} &=& {e\atop\kd{lvar1}}
\label{br8}\\
\left \{\;\int_{\cal R} d^np (\tilde{\omega}_{p}, \eta_p)\;,\;
(\xi_q|e|\eta_{q'})\;\right \} &=& 
(\xi_q|e|\eta_{q'}).
\nonumber \\
\left \{ {\pic{lvar6}{13}{-10}{2}{2}},{e\atop\kd{lvar1}} \right \} &=& 
{e\atop\kd{lvar1}}
\label{br9}
\end{eqnarray}
Here in (\ref{br8}) we assume $q$ is
contained in $\cal R$, and in (\ref{br9}) we assume $q'$
is contained in $\cal R$; otherwise the Poisson brackets vanish.

A nice feature of the graphical notation is that it gives a simple
mnemonic for the Poisson bracket formulas (\ref{br1}-\ref{br9}) above.
Note that in every case, the terms appearing in the Poisson bracket of
two of our observables correspond to the ways of attaching them
together by filling little empty circles (i.e.\ momentum fields) of
one observable with dots (i.e.\ configurational fields) of the other,
in a manner compatible with the orientations of the paths.  (In
theories for which particles coincided with their antiparticles, we
would not include orientation arrows on the paths in our graphical
notation, and not require compatibility of orientations when attaching
paths together.  This would give precisely the extra terms in the
Poisson bracket relations that actually appear in such theories.)

We see that the observables we have introduced are closed with respect
to Poisson brackets: a Poisson bracket of any two our observables is a
linear combination of these observables.  Note, however, that the path
observables involving fermionic momenta depend on a rule assigning to
any two points $p,p'\in\Sigma$ a path $e_{pp'}$.  The rules that
appear on the right hand side of (\ref{br1})-(\ref{br7}) are of the
form $e_{pq}\circ f_{qq'}$, i.e., they are formed by composition from
the rules we started with, so if we fix a rule we do not obtain a
Poisson algebra.  Instead, we work with the `big' algebra 
generated by all observables of the form (\ref{lvar1}-\ref{lvar6}), 
allowing all possible choices of a rule $e_{pp'}$ and all possible
regions $\cal R$ (open sets whose closure is compact).  
We denote this algebra as $\cal B$.

We may summarize by saying that we have introduced a Poisson algebra
$\cal B$ of smooth functions on the classical phase space $T^\ast{\cal
A} \times T^\ast {\cal F} $.  This algebra is not closed under the
Hermitian conjugation operator $\dagger$.  However, it has a
distinguished subalgebra of observables preserved by the $\dagger$
operation, namely the real-linear combinations of products of the
observables (\ref{lvar5}-\ref{lvar6}).

Our aim now is to quantize the observables (\ref{lvar1}-\ref{lvar6}),
finding operators on some Hilbert space whose commutators mimic the
classical Poisson bracket relations.   We proceed in
several steps.  We first construct a kinematical Hilbert space.  Then
we describe a subspace of gauge invariant states in this space, which
can be thought of as a Hilbert space of solution of Gauss law
constraint, and represent the observables in $\B$ as operators on this
space.

\section{Quantum configuration space and kinematical Hilbert space}
\label{sec:5}

Recall that for diffeomorphism-invariant theories of connections, the
quantum configuration space was most efficiently obtained by
completing the algebra of cylinder functions to obtain a commutative
C*-algebra, and taking the spectrum of this C*-algebra.  We
would like to follow a similar strategy for our theories with
fermions.  However, in our case the Gelfand-Naimark theorem is not
applicable because of the presence of Grassmann-valued fields.
Another strategy is needed.

Recall that in the case of theories of connections, we could also
describe the quantum configuration space as a certain completion of
the classical configuration space.  We can implement this idea in our
case and construct the fermionic quantum configuration space $\f$ as a
completion of the classical configuration space ${\cal F}$.  Namely,
we define $\f$ to be the space of all (not necessarily smooth)
sections of the vector bundle $F$ over $\Sigma$.  The space $\cal F$
is dense in $\f$ in the topology of pointwise convergence.

Alternatively, one can describe $\f$ as the projective limit of
configuration spaces of certain fermionic systems with finitely many
degrees of freedom.  Namely, given any finite subset of points $V
\subset \Sigma$ we may consider the configuration space of a system of
fermions living at the points of $V$; this is the product
\[     {\cal F}_V =  \prod_{p \in V} F_p  \]
of copies of fibers of $F$, one for each point in $V$.  
Given finite subsets $V \subseteq V'$ of points in $\Sigma$ there is a
natural projection from ${\cal F}_V'$ to ${\cal F}_V$, and the 
projective limit of all these spaces ${\cal F}_V$ is $\f$.

We also wish to have a kinematical Hilbert space for the fermion
fields.  Although we shall denote this Hilbert space
by $L^2(\f)$, it is not a space of square-integrable
functions on $\f$ in the standard sense.  We use the notation
$L^2(\f)$ just to emphasize the similarity in notations between the
fermionic and the gauge fields degrees of freedom.  

The space $L^2(\f)$ can be constructed
in a variety of equivalent ways.  Perhaps the easiest is to use the
fact that the fibers $F_p$ are finite-dimensional Hilbert spaces and
let $L^2(\f)$ be the fermionic Fock space over the Hilbert space
direct sum $\bigoplus_{p \in \Sigma} F_p$.  This is the natural
Hilbert space completion of the exterior algebra 
\[      \Lambda (\bigoplus_{p \in \Sigma} F_p) .\]
Alternatively, we can think of
$L^2(\f)$ as the infinite `grounded' tensor product
\[         \bigotimes_{p \in \Sigma} \Lambda F_p  \]
of the fermionic Fock spaces $\Lambda F_p$, one for each point $p$ in
space.  (For an introduction to the mathematics of fermionic Fock
spaces and infinite grounded tensor products, see 
\cite{BSZ}.)

In the case of connection fields we can define the kinematical Hilbert
space as a completion of an algebra of cylinder functions.  An
analogous fact holds for fermion fields if we define a `cylinder
function' on ${\cal F}$ to be a smooth function on ${\cal F}$
depending only on the value of the fermion fields at finitely many
points of $\Sigma$.  The space of cylinder functions is thus
\[   \Lambda\bigl(\bigoplus_{p\in\Sigma} F^*_p \bigr) 
\subseteq \Lambda \cal F^*, \]
the exterior algebra over the algebraic direct sum
$\bigoplus_{p \in \Sigma} F^*_p$.  Since each $F_p$ is a Hilbert space, 
this space is naturally isomorphic to 
\[   \Lambda\bigl(\bigoplus_{p\in\Sigma} F_p\bigr) , \]
and by the above remarks we see that it can
be completed to obtain $L^2(\f)$.

Finally, we may also describe $L^2(\f)$ in the following physically appealing
way.   For each finite set $V$ of points of $\Sigma$ we may start
with the theory of fermions living at these points.   The configuration
space for this theory is the finite-dimensional space ${\cal F}_V$
described above.   Quantizing this system we obtain the fermionic
Fock space over ${\cal F}_V$, which is just the exterior algebra
$\Lambda {\cal F}_V$ equipped with its standard inner product.  
Given finite subsets $V \subset V'$ of points in $\Sigma$ there is
a natural projection from $\Lambda {\cal F}_{V'}$ to $\Lambda {\cal F}_V$,
and the projective limit of these Hilbert spaces is the kinematical
Hilbert space $L^2(\f)$.  

Putting together the quantum configuration spaces for the gauge field
and fermionic degrees of freedom, we obtain the quantum configuration
space $\a \times \f$ for the whole theory.  Similarly, the kinematical
Hilbert space for the full theory is $L^2(\a)\otimes L^2(\f)$, which
we denote by $\hfull$.  Note that cylinder functions on $\cal A$ are
dense in $L^2(\a)$, and cylinder functions on $\cal F$ are dense in
$L^2(\f)$.  Thus if we define the algebra of `cylinder functions' on
${\cal A} \times {\cal F}$ to be the tensor product of the algebras of
cylinder functions on $\cal A$ and cylinder functions on $\cal F$, it
follows immediately that cylinder functions on ${\cal A} \times {\cal
F}$ are dense in $\hfull$.

In simple terms, a cylinder function on ${\cal A} \times {\cal F}$ is
a wavefunction depending on the connection only via its holonomies
along finitely many paths in $\Sigma$ and on the fermionic fields only
via their values at finitely many points of $\Sigma$.  As we shall see in
the next section, cylinder functions may be thought of as states of
gauge theories coupled to fermions living on graphs in $\Sigma$ with
finitely many edges and vertices.  Cylinder functions play an
important role in the quantum theory, for it is natural to define
quantum operators first on the dense subspace of cylinder functions,
and then to extend them to the entire Hilbert space $\hfull$.

The Hilbert space $\hfull$ serves only as an auxiliary Hilbert space
of the theory, since its states are typically not invariant under
gauge transformations, diffeomorphisms of space, or time evolution.
To find physical states we must impose constraints corresponding to
invariance under these symmetries.  In the next section we describe a
basis of solutions to the Gauss law constrant, 
that is, gauge-invariant elements
of $\hfull$.

\section{Gauge-invariant states}
\label{sec:6}

There is a natural unitary representation of the group $\g$ of smooth
gauge transformations on the Hilbert space of quantum states $\hfull$.
In this section we describe the subspace of $\hfull$ consisting of
that are invariant under these gauge transformations.  We denote this
space by $\hinv$.  We also describe an orthonormal basis for 
$\hinv$ that is a generalization of the spin network basis of
$L^2(\ag)$.

Given any graph $\gamma$ analytically embedded in $\Sigma$, 
let $E$ be its set of edges and let $V$ be its set of 
vertices.  Any edge $e \in E$ begins at some vertex $s(e)$ called
its `source' and ends at some vertex $t(e)$ called its 
`target'.  We define the space of connections on $\gamma$ to be
\[         {\cal A}_\gamma = \prod_{e \in E} {\cal A}_e \]
where ${\cal A}_e$ is the space of transporters from $s(e)$ to
$t(e)$.  If we trivialize $P$ at all the vertices of $\gamma$
we can think of ${\cal A}_\gamma$ as a product of copies of the
gauge group $G$, one for each edge.    Similarly, we define the
classical configuration space for fermion fields on $\gamma$ to be
\[        {\cal F}_\gamma = \prod_{p \in V} F_p . \]
We define $L^2({\cal A}_\gamma)$ using normalized Haar measure on $G$, 
define $L^2({\cal F}_\gamma)$ to be the fermionic Fock space over
${\cal F}_\gamma$, and set
\[       L^2({\cal A}_\gamma \times {\cal F}_\gamma) = L^2({\cal A}_\gamma)
\tensor L^2({\cal F}_\gamma) .\]
We may think of this as a subspace of $L^2(\a \times \f)$.  

For any graph $\gamma$ there is a unitary representation of $\g$ on
$L^2({\cal A}_\gamma \times {\cal F}_\gamma)$, and we denote the
subspace consisting of gauge-invariant elements by $L^2(({\cal
A}_\gamma \times {\cal F}_\gamma)/\g)$.  As in the case of pure gauge
fields \cite{AL,B}, the union of these spaces as $\gamma$ ranges over
all graphs is dense in $\hinv$.  Thus to describe $\hinv$ it suffices
to describe these space for arbitrary graphs $\gamma$.

Trivializing $P$ at all the vertices of $\gamma$ we have
\begin{equation}
L^2({\cal A}_\gamma \times {\cal F}_\gamma) =
\bigotimes_{e\in E} L^2(G) \otimes \bigotimes_{v\in V} \Lambda F_v.
\label{1}
\end{equation}
Let us decompose the Grassmann algebra $\Lambda F_v$ at each vertex
as an orthogonal direct sum of irreducible representations of $G$:
$\Lambda F_v = \bigoplus_{\rho\in S} \rho$, where we have 
denoted by $S$ the list of irreducible representations appearing in
$\Lambda F_v$.  Similarly, the Peter-Weyl theorem says that
$L^2(G) \iso \bigoplus_{\rho\in {\rm Rep}(G)} \rho\otimes\rho^*$, where 
the set ${\rm Rep}(G)$ contains one irreducible continuous 
unitary representation of $G$ from each equivalence class, so that 
(\ref{1}) implies
\begin{equation}
L^2({\cal A}_\gamma \times {\cal F}_\gamma) =
\bigotimes_{e\in E} \bigl
( \bigoplus_{\rho\in {\rm Rep}(G)} \rho\otimes\rho^* \bigr)
\; \otimes\; \bigotimes_{v\in V} \bigl ( \bigoplus_{\rho\in S} \rho \bigr ).
\label{2}
\end{equation}

The right hand side of (\ref{2}) can be rewritten as follows:
\begin{equation}
L^2({\cal A}_\gamma \times {\cal F}_\gamma) =
\bigoplus_{\rho_e\in {\rm Rep}(G),\; \rho_v\in S} \quad\bigotimes_{v\in V}
\bigl ( \bigotimes_{t(e)=v} \rho_e^* \otimes \bigotimes_{s(e)=v} \rho_e 
\otimes \rho_v \bigr )
\end{equation}
The sum here runs over all labelings of edges $e$ by irreducible 
representations $\rho_e$ of the group $G$, and by all labelings of 
vertices by irreducible representations from the list $S$.
It follows that 
\begin{equation}
L^2(({\cal A}_\gamma \times {\cal F}_\gamma)/\g) =
\bigoplus_{\rho_e\in {\rm Rep}(G), \rho_v\in S} \bigotimes_v \,\,
{\rm Inv}\bigl (\bigl( 
\bigotimes_{t(e)=v} \rho_e^* \otimes \bigotimes_{s(e)=v} \rho_e 
\bigr) \otimes \rho_v \bigr )
\label{3}
\end{equation}
where ${\rm Inv}$ denotes the $G$-invariant subspace of the given
representation.  Note that ${\rm Inv}\bigl ( \bigotimes_{t(e)=v}
\rho_e^* \; \otimes \; \bigotimes_{s(e)=v} \rho_e \otimes \rho_v \bigr
) $ has a natural inner product, and is isomorphic to the space of
intertwining operators from $\bigotimes_{t(e)=v} \rho_e$ to
$(\bigotimes_{s(e)=v} \rho_e) \otimes \rho_v$.  We denote this space
of intertwining operators by
\[   {\rm Hom}\bigl (
\bigotimes_{t(e)=v} \rho_e, \bigl(\bigotimes_{s(e)=v} \rho_e \bigr)
\otimes \rho_v \bigr ). \]

As in the case of pure gauge fields, the above discussion also gives
an orthonormal basis of spin network states for $\hinv$.
Unlike spin networks for theories describing only gauge fields, the
spin networks for theories with fermions can contain edges with `open
ends'.  Such open ends represent fermionic degrees of freedom.  
We describe the spin network basis for $\hinv$ in the 
following theorem summarizing the results of this section:

\begin{theorem} The space $\hinv$ has an orthonormal basis of 
`fermionic spin network states', each such state being specified by 
a choice of:
\begin{enumerate}
\item a graph $\gamma$ analytically embedded in $\Sigma$,
\item a labelling of each edge $e$ of $\gamma$ by an irreducible 
representation $\rho_e \in {\rm Rep}(G)$, 
\item a labelling of each vertex $v$ of $\gamma$ by an irreducible
representation $\rho_v \in S$,
\item a labelling of each vertex $v$ of $\gamma$ by an intertwining operator 
\[     \iota_v \in 
{\rm Hom}\bigl (
\bigotimes_{t(e)=v} \rho_e,\bigl( \bigotimes_{s(e)=v} \rho_e\bigr) 
\otimes \rho_v \bigr ) \]
chosen from an orthonormal basis of such intertwining operators.  
\end{enumerate}
\end{theorem}

\noindent We call the data $\Gamma = (\gamma,\rho,\iota)$ a `fermionic spin
network', and denote the fermionic spin network state corresponding to
$\Gamma$ as above by $\Psi_\Gamma$.   

\section{Quantization of fermionic path observables}
\label{sec:7}

We have now constructed the kinematical Hilbert space $\hfull$ as a
completion of the space of cylinder functions on $\a \times \f$, and
found a spin network basis for the gauge-invariant subspace $\hinv$.
Our construction so far has nothing directly to do with the 
fermionic path observables introduced in Sec. \ref{sec:4}.  Thus, we
have to make sure that there is a way to quantize these observables
and obtain operators on $\hinv$.  In fact, we shall obtain 
operators on $\hfull$ that commute with gauge transformations, and
thus map $\hinv$ to itself.

There is an obvious way to promote the configurational path
observables into operators.  We represent the quantities
$(\xi|e|\eta)$ by multiplication operators $(\xi|\hat{e}|\eta)$ on the
space of cylinder functions on ${\cal A} \times {\cal F}$ as follows:
\begin{equation}
     (\xi|\hat{e}|\eta) \Psi = (\xi|e|\eta)\,\Psi 
\label{lvar'1}
\end{equation}
where $(\xi|e|\eta)$ on the right hand side is considered as a
cylinder function on ${\cal A} \times {\cal F}$.  Note that if
$\Psi$ is a cylindrical function so is $(\xi|e|\eta)\Psi$. 
One has, therefore, a well-defined action of $(\xi|\hat e|\eta)$ 
on the space of cylinder functions.   

The observables (\ref{lvar2}-\ref{lvar6}) involving fermionic momenta
can also be first promoted to operators on the space of cylinder
functions.  The natural way to do this is to replace the fermionic
momentum field $\tilde{\pi}$ in such observables by the corresponding
Berezin derivative $\delta/\delta\xi$.  Berezin derivatives \cite{Berezin}
are usually used in the context of field theories in Minkowski spacetime.
However, we can also make sense of this notion in the
context of diffeomorphism-invariant theories.  

We introduce Berezin derivatives
as operators on the space of cylinder functions as follows.
Recall that the algebra of cylinder functions
on $\cal F$ is the exterior algebra
\[    \Lambda \bigl (\bigoplus_{p \in \Sigma} F^*_p \bigr). \]
For any point $p$ choose an orthonormal basis 
$\xi^\alpha(p)$ of the subspace of $F^*_p$ corresponding to the representation 
$\rho \in I$ under which the field $\xi$ transforms.   Note that
$\xi^\alpha(p)$ is a cylinder function on $\cal F$, and the algebra
of cylinder functions on $\cal F$ is generated by functions of this form.
Thus, formally speaking, we may define the Berezin derivative operator
to be the superderivation (i.e., graded derivation) of the algebra
of cylinder functions such that
\begin{equation}
{\delta\over\delta \xi^\alpha(p)}\,\xi^\beta(p') =\delta_\alpha^\beta
\tilde{\delta}^n(p,p'),
\end{equation}
where $\delta_\alpha^\beta$ is the Kroneker delta. 
However, the right hand side is distributional, so strictly speaking
we should smear this equation with a `test function'.  
In other words, for any section $f$ of $P \times_G \rho$, we 
define
\[ \int_\Sigma d^np\, f^\beta (p) {\delta\over\delta \xi^\beta(p)} \]
to be the unique superderivation on the algebra of
cylinder functions such that 
\[ \left( \int_\Sigma d^np \, f^\beta(p) {\delta\over\delta \xi^\beta(p)} 
\right) \xi^\alpha(p') = f^\alpha(p').  \]
To describe the action of this operator on an arbitrary
cylinder function, let us introduce the interior product operator
$i[\xi]_\alpha(p)$, namely the adjoint of exterior multiplication
by $\xi^\alpha(p)$.  We then have
\[ \left( \int_\Sigma d^np\, f^\beta(p) {\delta\over\delta \xi^\beta(p)} 
\right) \Psi = \sum_{p\in\Sigma} f^\alpha(p)\, i[\xi]_\alpha(p) \Psi, \]
where the right hand side, for any cylinder function $\Psi$, contains 
only a finite number of non-zero terms. 
 
Using this definition of Berezin derivative one can proceed with
the quantization of observables (\ref{lvar2}-\ref{lvar4}).  However,
there is an operator ordering ambiguity that needs to be resolved in
this procedure.  Since the observables (\ref{lvar2}-\ref{lvar4}) involve
both fermion fields and their canonically conjugate momenta, which do
not anticommute, we must specify which acts first.  Note that one of
our classical quantities, namely $\int_{\cal R} d^np
(\xi_p,\tilde{\pi}_p)$, has a natural interpretation as the number of
particles of type $\xi$ within the region $\cal R$.  To obtain this
number operator when we quantize, we need to choose the operator
ordering in which the Berezin derivative corresponding to the momentum
field $\tilde{\pi}_p$ acts before the multiplication operator
corresponding to the configuration field $\xi_p$.  Let us therefore demand that
derivatives act before multiplication operators for all our
observables, which amounts to a `normal ordering' prescription.

Following this choice of operator ordering we can now quantize all our 
fermionic observables, obtaining operators on the space of cylinder 
functions.   More precisely, we define the quantum analogs of 
the observables (\ref{lvar2}-\ref{lvar4}) as follows:
\begin{equation}
\int_{\cal R} d^np'\, (\xi_p|\hat e_{pp'}|{\tilde{\pi}}_{p'}) =
\sum_{p'\in{\cal R}} \, \xi^\alpha(p)\, ({\cal P} \int_e A)_\alpha^\beta \,
i[\xi]_\beta(p')   
\label{lvar'2}
\end{equation}
\begin{equation}
\int_{\cal R} d^np' \,( \tilde{\omega}_{p'}|\hat e_{p'p}| \eta_p) =
\sum_{p'\in{\cal R}} \,  \eta_\alpha(p)\, ({\cal P} \int_e A)^\alpha_\beta\,
i[\eta]^\beta(p')
\label{lvar'3}
\end{equation}
\[
\int_{\cal R} d^np \int_{\cal R'} d^np'
({\tilde{\omega}}_p|\hat e_{pp'}|{\tilde{\pi}}_{p'}) = \]
\begin{equation}  
\sum_{p\in {\cal R}}  \sum_{p'\in {\cal R'}}  
i[\eta]^\alpha(p)\, ({\cal P} \int_e A)_\alpha^\beta\,
i[\xi]_\beta(p')
\label{lvar'4}
\end{equation}
\begin{equation}
\int_{\cal R} d^np\, (\xi_p, \tilde{\pi}_p) = 
\sum_{p\in{\cal R}} \xi^\alpha(p)\, i[\xi]_\alpha(p)
\label{lvar'5}
\end{equation}
\begin{equation}
\int_{\cal R} d^np \,(\tilde{\omega}_p, \eta_p) = 
\sum_{p\in {\cal R}} \eta_\alpha(p)\, i[\eta]^\alpha(p)  
\label{lvar'6}
\end{equation}

The operators (\ref{lvar'1}-\ref{lvar'6})
map cylinder functions to cylinder functions, and 
extend uniquely to bounded linear operators on $\hfull$.  Since
they commute with gauge transformations they may also be thought of as
operators on $\hinv$.  Finite linear combinations of fermionic
spin network states are dense in $\hinv$, so the actions of
these operators on $\hinv$ are determined by their actions on 
fermionic spin network states.  
On such states the action is simply described in our graphical
notation, which is very nice for explicit computations.  Let us give
several simple examples; generalization to arbitrary spin network
states is straightforward.  We consider the action of our operators on
a fermionic spin network state containing a single path, that is, a
cylinder function of the form $(\xi|e|\eta)$.  First, 
the fermionic path observables linear in momentum,
(\ref{lvar'3}-\ref{lvar'4}), act to compose the path $e$ with
a path $f$.  That is,
\begin{eqnarray}
{\hat{f}\atop\kd{lvar2}}\circ\left | {e\atop\kd{lvar1}} \right
\rangle  = \left | {f\circ e\atop\kd{lvar1}} \right\rangle \\
{\hat{f}\atop\kd{lvar3}}\circ\left | {e\atop\kd{lvar1}} \right
\rangle  = \left | {e \circ f\atop\kd{lvar1}} \right\rangle 
\end{eqnarray} 
if the relevant endpoint of $e$ lies within the region $\cal R$;
otherwise we get zero.  
The operators (\ref{lvar'5}-\ref{lvar'6}) on the same state
are given as follows:
\begin{eqnarray}
{\atop\kd{lvar5}}\circ\left | {e\atop\kd{lvar1}} \right
\rangle  = \left | {f \atop\kd{lvar1}} \right\rangle \\
{\atop\kd{lvar6}}\circ\left | {e\atop\kd{lvar1}} \right
\rangle  = \left | {f\atop\kd{lvar1}} \right\rangle 
\end{eqnarray}
if the relevant endpoint of $e$ lies within the region $\cal R$;
otherwise we get zero.  In general, these operators
count the number of the fermions (resp.\ antifermions) in the region
$\cal R$. One can easily see that all fermionic
spin network states are eigenstates of these operators. 

The operator corresponding to a configurational path observable
acts to create a path:
\begin{equation}
{\hat{f}\atop\kd{lvar1}}\circ\left | {e\atop\kd{lvar1}} \right
\rangle  = \left | {f\atop\kd{lvar1}},{e\atop\kd{lvar1}} 
\right\rangle 
\end{equation}
The result is a spin network state containing two paths.
The operator involving two momentum fields acts to close off a path:
\begin{equation}
{\hat{f}\atop\kd{lvar4}}\circ\left | {e\atop\kd{lvar1}} \right
\rangle  = \left | {f\circ e\atop\pic{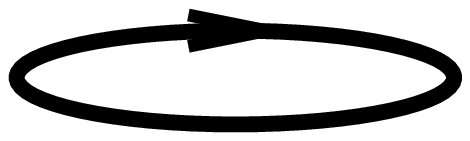}{10}{-3}{2}{2}} 
\right\rangle 
\end{equation}
if the endpoints of $e$ lie within the corresponding regions of
integration in the operator $\int_{\cal R} d^np \int_{\cal R'} d^np'
({\tilde{\omega}}_p|\hat e_{pp'}|{\tilde{\pi}}_{p'})$; otherwise we get
zero.  Note that the result here is the simple loop state so familiar from
the loop representation of pure gravity.  

In short, we can quantize our fermionic path observables, implementing
them as operators on $\hinv$.  As usual in quantization, we do
not obtain a true representation of the Poisson algebra $\cal B$,
in which Poisson brackets of arbitrary observables in 
$\cal B$ are mapped to commutators of the corresponding operators.
However, if we restore the factors of $\hbar$ which we have suppressed
above, and work with $\hbar \delta/\delta\xi^\alpha(p)$ where above
we use $\delta/\delta\xi^\alpha(p)$, we find that the commutators of
operators (\ref{lvar'1}-\ref{lvar'6}) match the Poisson brackets of
the observables (\ref{lvar1}-\ref{lvar6}) up to terms of order $\hbar^2$.

\section{Diffeomorphism-invariant states}
\label{sec:8}

The Hilbert space $\hinv$ of gauge-invariant states is still too large
to be the space of physical states of the theory. The group of
diffeomorphisms ${\rm Diff}(\Sigma)$ acts non-trivially on the
elements of $\hinv$.  To find diffeomorphism-invariant states one
should adopt the strategy described in Sec.\ \ref{sec:2}, namely the
`group averaging' technique of Ashtekar, Lewandowski, Marolf, Mour\~ao
and Thiemann \cite{ALMMT}.  Since the details of the construction are very
similar to those in the case of theories with pure connection degrees
of freedom, we only sketch the main steps here.

We look for diffeomorphism-invariant elements of the space ${\cal
C}^*$, the topological dual of the space $\cal C$ of gauge-invariant
cylindrical functionals.  We construct such solutions by averaging
fermionic spin network states, just as described in Sec.\
\ref{sec:2}.   Any state $\Psi \in \cal C$ that is a linear
combination of such states defines in this
way a diffeomorphism-invariant element $\overline \Psi$ of ${\cal
C}^*$.  There is an inner product on the space of such
diffeomorphism-invariant states given by
\[   \IP{\overline{\Psi}}{\overline{\Phi}} = 
\overline{\Psi}(\Phi). \]
Completing this space in this inner product, we obtain the Hilbert
space ${\cal H}_{\diff}$ of diffeomorphism-invariant states.  This
space has an orthogonal basis given by states of the form
$\overline \Psi_\Gamma$, where $\Gamma$ 
as we let $\Gamma$ range over a set of fermionic spin networks containing
one from each diffeomorphism equivalence class (again restricting
ourselves to graphs of type I).

\section{Conclusions}
\label{sec:D}

Having set up a kinematical framework for diffeomorphism-invariant
gauge theories coupled to fermions, and dealt with the Gauss law and
diffeomorphism constraints, there are still some interesting questions
concerning superselection sectors to deal with before moving on to the
much deeper problem of the Hamiltonian constraint.  In general one can
obtain superselection sectors (invariant subspaces) for the action of
an algebra on a Hilbert space by studying the orbits of various states
under the action of this algebra.  Let $\hat{\cal B}$ denote the algebra of
operators generated by the quantized fermionic path observables
(\ref{lvar'1}-\ref{lvar'6}) of Sec.\ (\ref{sec:7}).   It is especially
interesting to consider the orbit of the `empty state' under this algebra.
By the `empty state' we mean the vector in $\hinv$ given by
\[  1 \tensor 1 \in L^2(\A) \tensor \Lambda (\bigoplus_{p \in \Sigma} F_p) 
.\]
This is the fermionic spin network state corresponding to a graph 
with no vertices and no edges.  In general the empty state is not a 
cyclic vector for $\hat{\cal \B}$; i.e., the orbit $\hat{\cal B}(1 \tensor 1)$
is not dense in $\hinv$.  Thus $\hinv$ has nontrivial superselection sectors.  

There are various reasons why the empty state may not be a cyclic
vector.  First, there may be a nontrivial subgroup $H \subseteq G$
that acts trivially on all the representations in our list $I$
corresponding to the fermions in our theory.  If this occurs, the
empty state will never be a cyclic vector for $\hat{\cal \B}$.  To see
this, note that $H$ is a closed normal subgroup of $G$ so $G' = G/H$
is a compact connected Lie group.  We may naturally construct from $P$
a principal $G'$-bundle $P'$ and think of $F$ as associated to $P'$.
Letting ${\cal G}'$ denote the group of gauge transformations of $P'$,
and letting ${\cal A}'$ denote the space of smooth connections on
$P'$, one may check that $L^2({\cal A}' \times {\cal F}')/{\cal G}')$
is in a natural way a proper closed subspace of $\hinv$.  One may also
check that the orbit of the empty state lies in $L^2({\cal A}' \times
{\cal F}')/{\cal G}')$, so the empty state is not a cyclic vector.

It is easy to get around this problem, though, because in this
situation there is no harm in replacing the gauge group $G$ of our
theory by $G'$ and taking $L^2({\cal A}' \times {\cal F}')/{\cal G}')$
as our space of gauge-invariant states.  For example, consider the
gravitational and electromagnetic fields coupled to a charged
fermionic field as in Example B of Sec.\ \ref{sec:3}.  The particle in
this theory transform according to the spin $1/2$, charge $1$
representation of $G=\SU(2) \times \U(1)$, while its antiparticle
transforms according to the spin $1/2$, charge $-1$ representation.
The group $H$ thus consists of the elements $\pm (1,1)$, so
there is no harm in replacing $G$ by the group $G/H = \U(2)$.

Suppose therefore that this problem does not occur: only the identity
element of $G$ acts trivially on all the representations in the list
$I$.  The empty state may still fail to be a cyclic vector.  In fact,
this always occurs if $G = \SU(n)$ for odd values of $n > 2$ when the
list $I$ consists only of the fundamental representation and its dual.
The reason is that the $n$th exterior power of the fundamental
representation is the trival representation, so one may construct a
gauge-invariant quantity of order $n$ in the fermion field.  Applying
this to the empty state gives a fermionic spin network state living on
the graph with one vertex and no edges.  This element $\hinv$ cannot
be in the orbit of the empty state because the operators
(\ref{lvar'1}-\ref{lvar'6}) always change the total fermion number by
an even integer.  

However, the question of whether the empty state is a cyclic vector
remains open for the interesting cases of $G = \SU(2)$ and $G = \SU(2)
\times \U(1)$.  Another interesting open question is as follows.  At
least for $G = \U(n)$ and $G = \SU(n)$ one knows that products of
Wilson loop operators labelled by the fundamental representation span
$L^2(\a/\g)$.  As noted in Sec.\ \ref{sec:7}, such Wilson loops can be
expressed as products of fermionic path observables.  Therefore in
theories with these gauge groups and fermions tranforming under the
fundamental representation one knows that the closure of the orbit of
the empty state at least contains $L^2(\a/\g) \subseteq \hinv$.  For
what other groups and choices of representations does this result
hold?

\section{Acknowledgements}

We would like to thank H.\ Morales-T\'ecotl and Y.\ Shtanov for many
stimulating discussions when the basic ideas behind this work were
being developed.  K.\ K.\ also thanks A.\ Ashtekar for a useful
discussion.  We are grateful to the Erwin Schr\"odinger Institute for
Mathematical Physics in Vienna for their hospitality while much of
this paper was written, and especially to Peter Aichelburg and Abhay
Ashtekar for running the Workshop on Mathematical Problems of Quantum
Gravity held there.  J.\ B.\ also thanks the Center for Gravitational
Physics and Geometry for their hospitality while this paper was being
completed.  K.\ K.\ was supported, in part, by the International Soros
Science Education Program (ISSEP) through grant No.\ PSU062052, and this
research was also supported in part by NSF grant PHY95-14240.

\end{document}